\documentstyle[prl,aps,epsfig]{revtex}
\begin{document}
\draft
\title{
Derivation of a Quark Confinement Potential from QCD}
\author{Yongkyu Ko \footnote{e.mail : yongkyu@phya.yonsei.ac.kr}}
\address{
Department of Physics, Yonsei University, Seoul 120-749, Korea}
\date{\today}
\maketitle
\begin{abstract}
A quark confinement potential is derived from QCD Lagrangian under the two
assumptions in the energy region relevant to quark confinement. 
The possibility of perturbation in QCD is assumed, and the momentum of the 
gluon transfered by the quark is larger than the mass of the internal quark, 
so that the expansion of ${m^2 /{q^2}}$ is assumed to be reliable.
This potential comes from the one loop diagrams in the expansion of 
$m^2 /{q^2}$, especially here, the vacuum polarization diagram is examined.  
This explains the quark confinement mechanism on the basis of QCD and the
Regge trajectories which is governed by a linear potential.
\end{abstract}
\vspace{1cm}
Quarks are known as the constituents of the nucleon and the meson, and their
interaction is believed to be explained by quantum chromodynamics (QCD).
However there is no direct experimental evidence of detecting an isolated 
single quark.  So it is said that quarks are always confined in the hadron.  
QCD tells us that quarks behave as weakly interacting particles in high 
energy states, which is referred to as ``asymptotic freedom''.  In this 
energy states, perturbative calculations are possible so that perturbative 
QCD has been actively researched \cite{Field}.  In order to motivate this 
letter, it is useful to cite {\it ``If indeed the quarks behave 
approximately like free particles and their masses are rather small, the 
critical question, then, is why don't we see free quarks in the final state?  
This is the well-known problem of quark confinement.''}\cite{Lee}.   In low 
energy physics, however, perturbation is questionable because of the behaver 
of the larger coupling constant of QCD.  So quark degree of freedom doesn't 
appear manifest.  It is possible to describe nuclear phenomena
with the nucleon and the meson, i.e., meson theories instead of quarks.  
In order to describe low energy nuclear phenomena with quarks, several 
potential models are used such as bag models, harmonic oscillator potential, 
linear confinement potential model and so on \cite{Bhaduri,Perkins}.  Another 
kind of model is the string model from which a linear confinement potential is 
derived naturally, but in which the concept of a particle should be changed 
into a string \cite{Perkins}.   
All those potential models have their own quark confinement conditions.  

If QCD is a true theory for the interaction of quarks and gluons, why isn't 
the quark confinement phenomenon explained well with it?  The hint to this 
question is the Lamb shift.  With this hint, it is shown that QCD has 
a certain confinement potential which can be derived from QCD Lagrangian 
only.  Consequently, this potential gives the reason why quarks are always 
confined in the hadron.  Quark confinement 
is explained by the one loop effect of QCD, as if the Lamb shift is done by 
that of quantum electrodynamics (QED).  The essential difference between QCD 
and QED in the two phenomena is the magnitude of the coupling constants 
rather than the difference between the Abelian and non-Abelian gauge 
theories.  In QED, a hydrogen atom is the well-known bound state of a 
proton and an electron.  The coupling constant of QED is so small compared 
to that of QCD that the size of the hydrogen atom should be much larger than 
that of the nucleon which is a bound state of quarks.  The larger coupling 
constant will cause the larger momentum of the gluon which is transfered by 
the quark in order to localize quarks in the size of a nucleon.  

In the calculation of the Lamb shift, the momentum of the photon 
transfered by the electron is so small that the one loop
diagrams are calculated in the 
expansion of $q^2 / m^2$, and the result is the famous splitting of 
the 2p and 2s states of the energy levels of a hydrogen atom.  This 
splitting is generated by the following effective 
potential\cite{Ryder,Bjorken}:
\begin{equation}
V(r)= - {\alpha \over r} - {4 \alpha^2 \over 15} \delta(r)
\end{equation} 
in configuration space.  In momentum space, this potential corresponds to the
propagator of the photon which is modified by the one loop effect, i.e., 
the vacuum polarization:
\begin{equation}
D_{\mu \nu} =  {(g^{\mu \nu} - q^{\mu} q^{\nu}) \over q^2} 
(1- {e^2 \over{60 \pi^2}} {q^2 \over m^2}) .
\end{equation}
However what is calculated in such an expansion in QCD may be inappropriate 
to explain the hadronic phenomena, because quarks
are interacting  so strongly with each other.  
The quarks confined in the hadron may not be in low energy-momentum states 
because of the large coupling constant which will produce a large momentum 
transfer.  
So the basic assumptions of this letter are that the quarks inside a hadron 
are so relativistic that perturbation is meaningful and the expansion of the 
inverse momentum of the gluon, that is, $m^2 / q^2$ is reasonable rather 
than $q^2 / m^2$.

\begin{figure}
\centerline{\epsfig{file=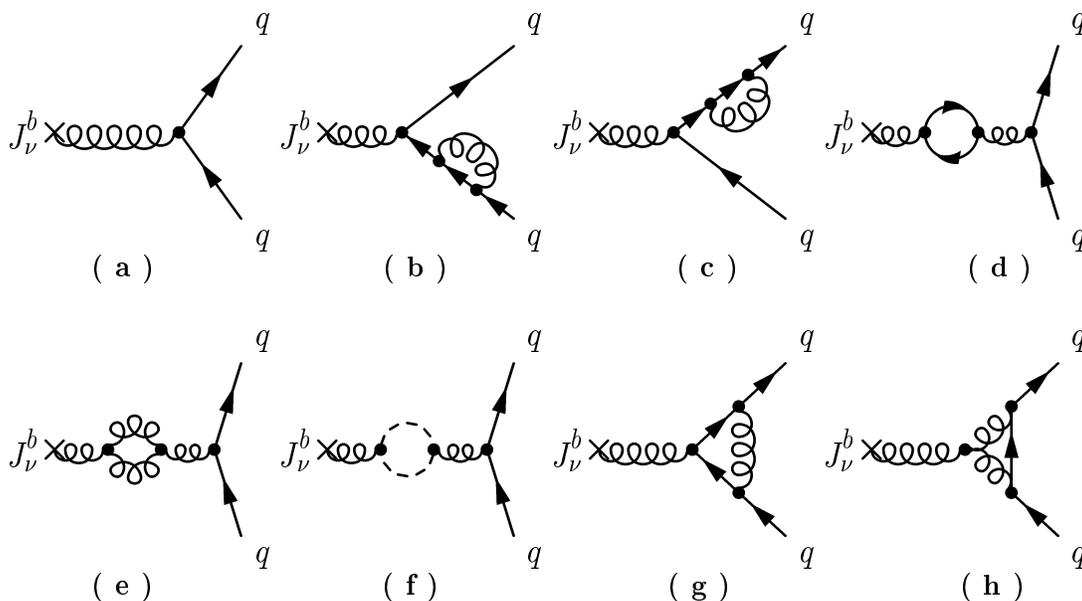, width=0.8\columnwidth}}
\caption{The Feynman diagrams at tree and one loop levels to contribute 
to the effective potential.  The diagrams for the counter terms are not shown.}
\label{feynman}
\end{figure}

The QCD Lagrangian is
\begin{eqnarray}
& &{\cal L} = i \bar{\psi } ( \partial_{\mu}-ig T^a A^a_{\mu} ) 
   \gamma^{\mu} \psi - m \bar{\psi} \psi - {1 \over 4} F^a_{\mu \nu} 
   F^{a \mu \nu},\nonumber \\
& &F^a_{\mu \nu} = \partial_{\mu} A^a_{\nu} - \partial_{\nu} A^a_{\mu} 
  + g f^{abc} A^b_{\mu} A^c_{\nu},
\end{eqnarray}
where the index $a$ runs 1,2,....,8. 
The effective potential to one loop order can be calculated from the Feynman 
diagrams of Fig. \ref{feynman}.  Those diagrams contribute to the following 
transition amplitude:
\begin{equation}
{\cal M}= g^2 \bar{u}(p_2) \{ \gamma_{\mu} F_1(q^2)
  +{ \kappa \over{2 m}}F_2(q^2) \sigma_{\mu \alpha} q^{\alpha} \} 
  T^a u(p_1) D^{\mu \nu}_{a b} J^b_{\nu},
\end{equation} 
where $J^b_{\nu}$ means the quark current which consists of the bispinor of an 
antiparticle for the meson or the composite of the bispinors of two quarks and 
gluons for the nucleon.  This amplitude is not a direct observable in the case 
of a bound state but an analogy of the amplitude of a scattering process.  
As the purpose of this letter is to show a qualitative confinement potential 
form QCD, the vacuum polarization diagram is considered only.  
It is calculated as 
\begin{eqnarray}
\Pi^{ab}_{\mu \nu}(q^2) &=& i g^2 \mu^{4-d} Tr(T^a T^b) 
  \int {d^dk \over{(2 \pi)^d}} \gamma_{\mu} {1 \over{ \not{\!q} 
  + \not{\!k} -m}} \gamma_{\nu}{1 \over{ \not{\!k} -m}}  \nonumber \\
&=& - {g^2 n_f \delta_{ab}\over{12 \pi^2}} ( q^2 g_{\mu \nu}-q_{\mu} 
  q_{\nu}) \{{1 \over{\epsilon}}-{\gamma \over 2}-{1 \over 2} 
  \text{log}({- m^2 \over{4 \pi \mu^2}}) \} + \Pi_{\text{finite}}(q^2),
  \label{cvacuum}
\end{eqnarray}
where $n_f$ is the number of flavors to contribute to the loop of the 
quark \cite{Ryder}.  The infinite part in the above result contributes to the 
renormalization of the wave function of the gluon.  The remaining finite 
part is calculated as
\begin{eqnarray}
& &\Pi_{\text{finite}}(q^2) = {g^2 n_f \delta_{ab}\over{4 \pi^2}} 
  (q^2 g_{\mu \nu}-q_{\mu} q_{\nu}) \int^1_0 dx x(1-x) \text{log}
  ({x(1-x)q^2-m^2 \over{-m^2}}) \nonumber \\
&=& {g^2 n_f \delta_{ab}\over{4 \pi^2}} (q^2 g_{\mu \nu}-q_{\mu} q_{\nu}) 
  \{- {5 \over 18} - {2m^2 \over 3q^2} 
+ {1 \over 6}(1 + {2 m^2 \over q^2}) \sqrt{1 - {4 m^2 \over q^2}} 
  \text{log}({\sqrt{1 - {4 m^2 \over q^2}} +1 
  \over {\sqrt{1 - {4 m^2 \over q^2}} -1 }})  \}.\label{vacuum}
\end{eqnarray}
If it is expanded in $-{4 m^2 /{q^2}}$ which is called the expansion 
of ${m^2 /{q^2}}$ throughout this letter, it is easy to calculate the effective
potentials.  This vacuum polarization modifies the gluon propagator as
\begin{eqnarray}
D^{ab}_{\mu \nu}(q^2) &=& {\delta^{ab} ( g^{\mu \nu}-q^{\mu} q^{\nu}) \over q^2}
+ {\delta^{ac}( g^{\mu \rho}-q^{\mu} q^{\rho}) \over q^2} \Pi^{cd}_{\rho \sigma}
 {\delta^{bd} (g^{\sigma \nu}-q^{\sigma} q^{\nu}) \over q^2} 
  +\cdot \cdot \cdot \cdot \cdot \cdot \nonumber \\
&=& {\delta^{ab} (g^{\mu \nu}-q^{\mu} q^{\nu}) \over q^2} 
  [ 1+ {g^2 n_f \over{4 \pi^2}} \{- {5 \over 18} - {1 \over 6} 
  \text{log}(-{m^2 \over q^2}) - {m^2 \over q^2} \nonumber \\
&-& {1 \over 2} ({m^2
 \over q^2})^2 +({m^2 \over q^2})^2 \text{log} (-{m^2 \over q^2}) 
 + \cdot \cdot \cdot \cdot \cdot \cdot \}
 + \cdot \cdot \cdot \cdot \cdot \cdot ], \label{gluonp}
\end{eqnarray}
where the coupling constant $g$ is renormalized one from now on, since the 
infinite part of the vacuum polarization is canceled by the suitable counter 
term which comes from the redefinition of the physical parameters, i.e., 
wavefunctions, $m,g$ and so on.  In the center of mass frame of a quark and 
the remaining part, the four momentum of the gluon transfered by the quark 
is reduced to a three momentum, because the energy of the quark before the
interaction with the gluon is exactly equal to that after the interaction.
Now the Fourier transforms of each term are 
\begin{eqnarray}
&{}& \int {d^3q \over {(2 \pi)^3}}{1 \over {{\bf q}^2}}e^{i{\bf q \cdot r}} 
  = {1 \over{ 4 \pi r}},\nonumber \\
&{}& \int {d^3q \over {(2 \pi)^3}} {1 \over{{\bf q}^2}}\text{log}({\bf q}^2)
  e^{i{\bf q \cdot r}} = - {1 \over{ 2 \pi r}}\text{log}(\gamma r),\nonumber \\
&{}& \int {d^3q \over {(2 \pi)^3}}{1 \over{\bf q}^4} e^{i{\bf q \cdot r}} 
  = - {1 \over{8 \pi}} r,\nonumber \\
&{}& \int {d^3q \over {(2 \pi)^3}}{1 \over{\bf q}^6} e^{i{\bf q \cdot r}} 
  =  {1 \over{4! 4 \pi}} r^3, \nonumber \\
&{}& \int {d^3q \over {(2 \pi)^3}}{1 \over{\bf q}^6} \text{log}({\bf q}^2) 
  e^{i{\bf q \cdot r}} =  {1 \over{4! 2 \pi}} \{ {25 \over{12}} - \text{log}
  (\gamma r) \} r^3,
\end{eqnarray}
where the table of integrals of Ref. \cite{Gradshteyn} is used and
$\gamma$ is the Euler constant in the second Fourier transform.  The first 
one is the usual Coulomb potential as encountered in QED.  The second one 
is known as the Uehling potential\cite{Uehling}.  The third one is the linear 
potential which confines quarks in the hadron forever.  The last two equations 
are only for reference, which give higher order confinement potentials.  
Alternatively, the linear confinement potential can be also estimated from 
the exact equation (\ref{vacuum}) by ignoring the last term which behaves 
approximately like ${1 \over{\bf q}^2}\log({\bf q}^2)$ at large momentum 
in the gluon propagator, so one can check that the integration interval in the 
Fourier transformations and higher power terms in Eq. (\ref{gluonp}) do not 
matter.
From the Fourier transform of the propagator of the gluon, the effective 
potential by which a quark is influenced is calculated as
\begin{equation}
V(r)= - c_F{\alpha_s \over r} + {5 c_F n_f \over{18 \pi}}{\alpha_s^2 \over r}
  + {c_Fn_f \over{3 \pi}}{\alpha_s^2 \over r}\text{log} (\gamma m r) 
  + {c_F n_f \over{ 2 \pi}}m^2 \alpha_s^2 r 
  + \cdot \cdot \cdot \cdot \cdot, \label{potential}
\end{equation}
where $c_F$ is the color factor which is ${4 \over 3}$ for the meson 
and ${2 \over 3}$ for the baryon and $\alpha_s = {g^2 \over{4 \pi}}$ 
\cite{Halzen}.  The non-Abelian nature of QCD gives only a constant, such as
the color factor from the trace in Eq. (\ref{cvacuum}).  The diagrams
Fig. 1-(e) and (f) which are quite different from QED do not give
any confinement potential in the gluon propagator but give only Uehling 
potentials because there is no internal quark line.  The complete
effective potential may be needed to consider further through out the 
remaining diagrams of Fig. 1 and bremsstrahlung diagrams.

There is the string model that has such a linear potential \cite{Perkins}.
This model explains a linear dependence of the angular momentum of the states 
of highest angular momentum, on the square of the mass, namely,
$ J = \alpha' E^2 + \text{constant}$.  This relation holds for the case of 
a constant energy density $k$ of a rotating string, that is, to a potential 
of the form $V = k r$.  The value of $k$ is obtained from the slope $\alpha'$ 
with the relation $ \alpha' = {1 / {2 \pi k}}$, which is derived in 
Ref. \cite{Perkins}.
Inserting the observed value $\alpha' = 0.93~ \text{GeV}^2$ from the plot of 
spin $J$ against mass squared for baryon resonances of the $\Delta$ 
family ( s = 0, I = $3 \over2$ ) and the $\Lambda$ family ( s = -1, I = 0 ), 
the coefficient $k$ is $0.85~ \text{GeV fm}^{-1}$.  Comparing it 
with Eq. (\ref{potential}), then the coefficient $k$ is matched to the result
of this work as $ k = {c_F \alpha_s^2 \over{2 \pi}} \sum_{f=u,d,s,...}m^2_f$,
where the mass squared is summed up to the quark satisfying the 
condition $-{4 m^2_f /{q^2}}<1$.  Therefore the Regge trajectories can be 
regarded as a strong experimental evidence of the linear confinement potential.

There is a theory that has a large coupling constant but not a confinement 
potential.  It is the very N-N potential in nuclear physics which is produced 
by the force mediated by pions, vector mesons and so on by coupling to nucleons.  The coupling constant 
of $g^2_{\pi NN} / {(4 \pi)}$ is about 14.3 in the 
literature \cite{Weise}.  The pion mass is the reason why there is no such
a potential.  The pion propagator corrected by the one loop effect 
in the expansion of $m^2 /{q^2}$ is
\begin{eqnarray}
D(q^2) &=& {1 \over {q^2-m^2}} +{1 \over {q^2-m^2}} \Sigma {1 \over {q^2-m^2}} 
  + \cdot \cdot \cdot \cdot \cdot \cdot \nonumber\\
 &=& {1 \over {q^2-m^2}} [ 1 + {g^2 \over {4 \pi^2}} 
  \{ C_1 + C_2 \text{log}(q^2) + C_3 {m^2 \over {q^2}} 
  + \cdot \cdot \cdot \cdot \cdot \cdot \}
  + \cdot \cdot \cdot \cdot \cdot \cdot ], 
\end{eqnarray}
where the free propagator should not be expanded because of renormalizability.  
The Fourier transform of the term which has the coefficient $C_3$ and is 
expected to give a confinement potential like Eq. (\ref{gluonp}) still has 
an exponential decay factor due to the mass of the pion as followings:
\begin{equation}
\int {d^3q \over {(2 \pi)^3}} {m^2 \over {{\bf q}^2 ({\bf q}^2+m^2)}} 
  e^{i{\bf q \cdot r}} = {1 \over{4 \pi r}}(1- e^{-mr}).
\end{equation}
The higher order terms of ${m^2 /{ q^2}}$ also do not give a confinement
potential, which can be proved by differentiating the above Fourier transform 
twice with respect to $r$.  
Therefore the N-N potential has no confinement potential, though it has a 
large coupling constant.

However there remains still a question: how large coupling constant produces 
such a confining potential?  Since such a potential can be also derived in QED, 
if the possibility of the expansion of ${m^2 /{ q^2}}$ is forced to assume,  
it is necessary to justify the assumptions of this letter.  The running
coupling constant of QCD is derived in the perturbative region, so in the 
non-perturbative region, it is useful to use that the velocity of the 
electron $v$ is equal to the coupling constant $\alpha$ in the Bohr model.  
Using the relativistic energy momentum relation
$p^2_H=M^2_H=(p_e+p_p)^2$ in the center of mass frame, the identity is 
calculated as
\begin{equation}
v_e^{non}={|{\bf p}_e| \over{\mu_e}}= \alpha=
  {(m_e + m_p) \sqrt{|(M_H^2-m_e^2-m_p^2)^2-4m_e^2m_p^2}| \over{2 M_H m_e m_p}} 
  \approx (137.02)^{-1},\label{velocity}
\end{equation}
where the subscripts $H,e$ and $p$ mean the hydrogen, electron and proton, 
respectively, and $\mu_e$ is the reduced mass of the electron.
Solving this equation for the mass of the hydrogen, and it is expanded with
respect to the coupling constant and  $m_e/m_p$ as
\begin{equation}
M_H=m_p + m_e - {1 \over 2} \alpha^2 \mu_e - {1 \over 8} \alpha^4 \mu_e 
+ {3\over8}\alpha^4 \mu_e {m_e \over{m_p}} - {1\over16} \alpha^6 \mu_e + \cdot
 \cdot \cdot \cdot \cdot \cdot,
\end{equation}
where the fine structure term ($\alpha^4$), which consists of the spin orbit 
coupling and relativistic corrections, can be derived correctly for the ground 
state.  The hyperfine structure term appears less reliable, because the 
spin-spin interaction is not applied to the equation.
The equation is also applied to the positronium as
\begin{equation}
M_{posi}=2 m_e - {1 \over 4} \alpha^2 m_e - {1 \over 64} \alpha^4 m_e + \cdot
 \cdot \cdot \cdot \cdot \cdot.
\end{equation}
Calculating Eq. (\ref{velocity}) for the deuteron, the result is 0.097 
which is the order of the pseudovector coupling constant ${f^2 /{(4 \pi)}}$  
for the usual definition ${f /{m_{\pi}}}={g_{\pi NN}/{(2 M)}}$ \cite{Weise}.  
The ratio of this result to the usual value 0.079 is similar to the axial 
vector coupling constant $g_A$.  

From the above reasonable results, the strong coupling constant can be 
estimated to a certain extent as
\begin{equation}
\alpha_s = {\sqrt{| M_{\pi^0}^2 - 4 m_q^2|} \over{m_q}} 
= {2|{\bf p}_q| \over{m_q}},\label{salpha}
\end{equation}
where $\pi_0$ has a quark and an anti-quark with the same mass.
Since the relativistic velocity is less than 1, the lightest quark mass is 
allowed to be less than 95.4 MeV just as $ v_q^{rel}= {|{\bf p}_q| /{E_q}} 
= { \sqrt{|M_{\pi^0}^2 - 4 m_q^2 | } /{M_{\pi^0}}} < 1$.
Hence the coupling constant should be estimated from the current quark masses.
Now the momentum of the gluon is estimated as
\begin{equation}
q^2=-4{\bf p}_q^2 \sin^2({\theta /2})=-{m_q^2\alpha_s^2 
  \sin^2({\theta /2})},\label{mgluon}
\end{equation}
in the center of mass frame.  Though this equation is defined for scattering
processes, if it is applied to QED and compared with the Bohr radius, then
the relation $\sin^2{(\theta /2)}={1\over4}$ can be analogized for the ground 
state 
($|{\bf q}| \approx <\psi_{100}|1/r|\psi_{100}>={1 /{a_{Bohr}}}=m_e \alpha$).

For a numerical example, the coupling constant $\alpha_s$ is estimated as 
13.35 for the typical current quark mass $m_q=10$ MeV and then the gluon
momentum $q=66.74$ MeV from Eqs. (\ref{salpha}) and (\ref{mgluon}).
The gluon momentum transfered by an excited quark can be estimated simply as 
201.72 MeV (66.74+134.98) and 836.74 MeV, respectively, at the threshold 
energies of $\pi_0$ and $\rho_0$.  The running
coupling constant $\alpha_s(q^2)={4 \pi \over{\beta \log(q^2/\Lambda^2)}}$
can be extrapolated by using unknown parameters 
$\beta,\Lambda$ for $\alpha_s(66.74~ \text{MeV})=13.35$
and $\alpha_s(m_{\tau})=0.35$ which is the measured coupling constant at the
lowest mass scale \cite{Review,European}.  
The coupling constants can be estimated as $\alpha_s(201.72~ \text{MeV})=0.988$
and $\alpha_s(836.74~ \text{MeV})=0.451$,
and the expansion parameters of ${m^2 /{q^2}}$ are 0.0025 and 0.00014, 
respectively, in the energy region of the $q \bar q$ creation.
Therefore perturbation with respect to the $\alpha_s$ and the expansion of
${m^2 /{q^2}}$ are reliable simultaneously in the energy region relevant
to quark confinement.  So, quarks are confined in the hadron by the strong
coupling in the non-perturbative region and by the linear confinement potential
in the perturbative region.  If the assumptions of this letter hold for the 
lightest meson, they also hold for all the other hadrons, because the mass of
the internal quark loop starts from the lightest quark.  In QED, on the 
contrary, the inner most electron of the most heavy atom can not transfer a 
momentum lager than $2 m_e$, because the sum of the binding energy 
($13.6 \times 120^2 ~\text{eV}=0.196~ \text{MeV}$) for the atomic number 
$Z=120$ and the momentum of the photon (0.448 MeV) is estimated as 0.644 MeV.
Therefore one can not see any confinement potential in QED, because the 
expansion of ${m^2 /{q^2}}$ is not allowed.

As a result of the view of this letter, quark masses are not direct
observables but parameters to be obtained from the hadronic spectra by
solving the Dirac equation.  So quark masses are unknown exactly.
The current quark masses support the assumptions of this letter and the mass
of the light mesons quite well rather than the constituent quark masses.
If the quarks are bounded by the usual
Coulomb potential, then the mass of the light meson is less than the sum of
the masses of its constituent two quarks because of their negative kinetic
energies.  However the quark are bounded by the confinement potential of
this letter, the mass of the meson is larger than the sum of the masses
of the two quarks because of their positive kinetic energies.

\begin{center}
\acknowledgments
This work was supported by the Korean Ministry of Education(Project no.
BSRI 97-2425).
\end{center}

\end{document}